# A Performance Comparison of Stability, Load-Balancing and Power-Aware Routing Protocols for Mobile Ad Hoc Networks

Natarajan Meghanathan[1] and Leslie C. Milton[2]

[1]Jackson State University, 1400 John Lynch St, Jackson, MS 39217, USA
natarajan.meghanathan@jsums.edu
[2]US Army Engineering Research and Development Center, Vicksburg, MS 39180, USA
Leslie.C.Milton@usace.army.mil

**Abstract**

The high-level contribution of this paper is a simulation-based detailed performance comparison of three different classes of routing protocols for mobile ad hoc networks: stability-based routing, power-aware routing and load-balanced routing. We choose the Flow-Oriented Routing protocol (FORP), the traffic interference based Load Balancing Routing (LBR) protocol and Min-Max Battery Cost Routing (MMBCR) as representatives of the stability-based routing, load-balancing and power-aware routing protocols respectively. Among the three routing protocols, FORP incurs the least number of route transitions; while LBR incurs the smallest hop count and lowest end-to-end delay per data packet. Energy consumed per node is the least for MMBCR, closely followed by LBR. MMBCR is the most fair in terms of node usage and hence it incurs the largest time for first node failure. FORP tends to repeatedly use nodes lying on the stable path and hence is the most unfair of the three routing protocols and it incurs the smallest value for the time of first node failure. As we measure the failure times of up to the first five nodes in the network, we observe that LBR incurs the maximum improvement in the lifetime of the nodes and MMBCR incurs the least improvement beyond the time of first node failure.

## 1. Introduction

A mobile ad hoc network (MANET) is a dynamic distributed system of wireless nodes that move independently and arbitrarily. The wireless medium is shared and the transmissions are prone to interference. MANET nodes operate with reduced battery charge and have limited transmission range. Hence, several routing protocols have been proposed for MANETs. Proactive routing protocols determine routes for every pair of nodes in the network, irrespective of the requirement. The reactive or on-demand routing protocols determine a route only when required, using a broadcast query-reply cycle. In dynamic scenarios, typical to that of MANETs, reactive on-demand routing protocols incur less overhead and exhibit better performance compared to the class of proactive routing protocols [1,2]. Hence, we restrict ourselves to the on-demand routing protocols in this paper.

Depending on the principle and/or the metric behind route selection, MANET routing protocols can be categorized into different classes: power-aware routing, load-balanced routing, minimum-hop/delay based routing, stability-based routing and etc. Examples of minimum-hop/delay based routing protocols include proactive protocols like the Destination Sequenced Distance Vector (DSDV) routing protocol [3] and the link-state based routing protocols (like Fish eye State Routing [4]) and on-demand protocols like the Dynamic Source Routing (DSR) [5], Ad hoc On-demand Distance Vector (AODV) routing [6], Location-aided Routing (LAR) [7] and the Temporally Ordered Routing Algorithm (TORA) [8]. Examples of stability-based routing protocols are the Flow-oriented Routing Protocol (FORP) [9], Associativity-based Routing (ABR) protocol [10] and the Route-lifetime Assessment Based Routing (RABR) protocol [11]. Similarly, several on-demand power-aware routing protocols (e.g., PCR [12], PSR [13], DEAR [14] and etc.) have been also proposed in the literature. But, these power-aware routing protocols are not designed to optimize the time of first node failure. Min-Max Battery Cost Routing (MMBCR) [15] is a power-aware routing algorithm proposed to maximize the time of first node failure as this routing algorithm always selects routes with the objective of maximizing the minimum residual battery power of a node in the route. MMBCR can be implemented on the top of any on-demand routing protocol like DSR, AODV and etc. The Load-balancing routing (LBR) [16] attempts to route data packets by circumventing congested paths and balances the traffic load to yield a lower end-to-end delay per data packet. LBR outperforms both DSR and AODV by yielding a higher packet delivery ratio and a lower end-to-end delay per data packet [16].

We refer to transmission power control as the technique of dynamically adjusting the transmission power of the sending node based on the distance to the intended receiving node of the packet [17]. Note that for clarity purposes, we denote the end nodes of a hop as the sender and receiver and refer to the end nodes of a path as the source and destination. In the scenario where there is no transmission power control, the transmission power per hop is fixed and is just based on the transmission range of the sender node. Using transmission power control, the transmission power spent to send a packet on a hop is a function of the distance between the sender (also called the transmitter) and receiver, which is less than or equal to the transmission range of the sender node. Transmission power control provides a couple of significant advantages: First, it helps to reduce the energy consumed in sending a packet from a source to destination across multiple hops. Second, it increases the bandwidth usage, because during a transmission/reception at a hop we freeze (from transmission and reception) only the nodes whose distance to the transmitter or the receiver is less than or equal to the distance between the transmitter and the receiver. As a result, the chances of packet collisions are also reduced.

To the best of our knowledge, we could not find any work that has compared the performance of the following three different categories of routing: power-aware routing protocols to maximize the time of first node failure, load-balanced routing and the stability-based routing protocols. In this work, we choose MMBCR (implemented on the top of DSR), LBR and FORP to be the representatives of the power-aware routing protocols to maximize the time of first node failure, load-balanced routing and the stability-based routing categories. We target peer-to-peer ad hoc networks such as the personal communication networks where wireless devices are available with individuals. Loss of connectivity to even one node is significant as any node can be come a source or destination. Accordingly, we are interested in measuring the time of first node failure. All nodes are to be treated equally and fairly in terms of node usage. We implement all these three routing protocols in ns-2 and study their performance with respect to several metrics, both in the absence and in the presence of power control. The performance metrics studied are: the number of route transitions, the average hop count per path, the end-to-end delay per data packet, the energy consumed per data packet, the energy consumed per node, fairness of node usage (in terms of the standard deviation of the energy consumed per node), the time of first node failure and the times of subsequent node failures (up to the 5$^{th}$ node failure).

The rest of the paper is organized as follows: Section 2 provides an overview of the FORP, LBR and MMBCR protocols. Section 3 describes the simulation environment and introduces the performance metrics measured. Section 4 illustrates the performance results for each metric and interprets them. Section 5 summarizes the performance results and concludes the paper.

## 2. Review of MANET Routing Protocols

In this section, we provide a brief overview of the stability-based Flow-Oriented Routing Protocol (FORP), load-balancing routing (LBR) protocol and the power-aware Min Max Battery Cost Routing (MMBCR) protocols. All the three routing protocols employ a flooding-based query-reply cycle to discover the routes in an on-demand fashion.

### *2.1 Flooding-based Query-Reply Cycle*

We describe here a generic flooding-based route discovery approach that we use to discover routes for the routing protocols studied in this paper. The route selection policies of the routing protocols are integrated with the route discovery approach. Whenever a source node $s$ has data to send and does not know about any route to a destination node $d$, the source node initiates flooding by propagating a Route-Request (RREQ) packet among its neighbours. The RREQ packets of successive route discoveries are uniquely identified by monotonically increasing sequence numbers. Each intermediate node upon receiving the RREQ packet will rebroadcast the packet if the node has not seen a RREQ packet with a sequence number greater than or equal to that in the current RREQ packet. Before forwarding the RREQ packet, the intermediate node inserts its own ID on the RREQ packet and updates the cost field for the upstream link on which the RREQ was received. The RREQ packet is then forwarded to the neighbours. The destination node receives RREQ packets along several paths and selects the path that best satisfies the route selection principles/ metric of the particular routing protocol in use. The destination sends a Route-Reply (RREP) packet on the reverse of the selected path so that the packet now travels from the destination back to the source. The destination node also includes the link-wise $s$-$d$ path information in the RREP packet. The intermediate nodes on the selected path will learn about their inclusion in the $s$-$d$ path after receiving the RREP packet. The source node learns about the $s$-$d$ path from the RREP packet and starts transmitting the data packet on the learnt $s$-$d$ path. If any RREQ is received along a path which is better (with respect to the route selection principle/ metric employed by the routing protocol) than the path already selected, another RREP would be sent on the better path discovered. When an intermediate node on an $s$-$d$ path cannot forward the data packet to a downstream node, the intermediate node sends a Route-Error (RERR) packet to the source node, which initiates a new flooding-based route discovery.

### *2.2 Flow-Oriented Routing Protocol (FORP)*

FORP [9] utilizes the mobility and location information of nodes to approximately predict the expiration time (LET) of a wireless link. The minimum of LET values of all wireless links on a path is termed as the route expiration time (RET). The route with the maximum RET value is selected. Each node is assumed to be able to predict the LET values of each of its links with neighboring nodes based on the information regarding the current position of the nodes, velocity, the direction of movement, and transmission range. FORP assumes the availability of location-update mechanisms like GPS (Global Positioning System) [21] to identify the location of nodes and also assumes that the clocks across all nodes are synchronized. Route discovery is similar to the flooding-based query-reply cycle described in Section 2.1, with the information propagated in the RREQ packet being the predicted LET of each link in a path.

Given the motion parameters of two neighbouring nodes, the duration of time the two nodes will remain neighbours can be predicted as follows: Let two nodes $i$ and $j$ be within the transmission range of each other. Let $(x_i, y_i)$ and $(x_j, y_j)$ be the co-ordinates of the mobile hosts $i$ and $j$ respectively. Let $v_i$, $v_j$ be the velocities and $\Theta_i$, $\Theta_j$, where ($0 \leq \Theta_i$, $\Theta_j < 2\pi$) indicate the direction of motion of nodes $i$ and $j$ respectively. The amount of time the two nodes $i$ and $j$ will stay connected, $D_{i-j}$, can be predicted using the following equation:

$$D_{i-j} = \frac{-(ab+cd) + \sqrt{(a^2+c^2)r^2 - (ad-bc)^2}}{a^2 + c^2} \quad (1)$$

where,
$a = v_i \cos\Theta_i - v_j \cos\Theta_j$; $b = x_i - x_j$; $c = v_i \sin\Theta_i - v_j \sin\Theta_j$; $d = y_i - y_j$

RREQ packets are propagated as described before, from the source node $s$ to the destination node $d$. The information recorded in this case by a node $j$ receiving a RREQ packet from a node $i$ is the predicted lifetime of the link $i$-$j$. The destination $d$ will receive several RREQ packets with the predicted link lifetimes in the paths traversed being listed. The residual expiration time of a path is the minimum of the predicted lifetimes of its constituent links. The $s$-$d$ path that has the maximum predicted residual lifetime is then selected. If more than one path has the same maximum predicted residual lifetime, the tie is broken by selecting the shortest of such paths.

### *2.3 Load-Balancing Routing (LBR) Protocol*

The LBR protocol [7] uses the concepts of "node activity" and "traffic interference" to select the best $s$-$d$ path that would encounter the minimum traffic load in transmission and minimum interference by neighbouring nodes. The activity of a node is defined as the number of active $s$-$d$ paths ($s$-$d$ paths that currently use the node as one of the intermediate forwarding nodes) the node is part of. The traffic interference at a node is the sum of all the activities of the neighbors of the node. For a given source $s$ and destination $d$, LBR chooses an $s$-$d$ path such that the sum of the traffic interferences and the activities of the intermediate forwarding nodes on the path is the minimum. The route selection metrics recorded in the RREQ packets are the activity and traffic interference of each of the intermediate forwarding nodes of the RREQ packet.

If $A_i$ is the number of active *s-d* paths in which node *i* is part of, the traffic interference at node *i* is $TI_i = \sum_{\forall j} A_j^i$, where *j* is a neighbouring node of node *i*. The cost of path *k* is defined as $C_k = \sum_{i \in k}(A_i + TI_i)$, where *i* is a node, other the source and destination, on path *k*. LBR chooses the *s-d* path that has the minimum cost value.

## 2.4 Min-Max Battery Cost Routing (MMBCR)

The residual battery charge of an *s-d* path is the minimum of the battery charges of the intermediate nodes of the path. The MMBCR algorithm [15] chooses the *s-d* path with the largest residual battery charge. The route selection metric recorded in an *s-d* path is the residual battery charge (available battery charge) of each of the intermediate nodes on the *s-d* path through which the RREQ packet got forwarded. If $B_i$ is the available residual battery charge of node *i*, then the residual battery charge of a path *k* is $Min_{i \in k}(B_i)$, the minimum of the residual battery charge of all the constituent intermediate nodes, other than the source and destination, of the path. From all the *s-d* paths learnt, the destination chooses the path that has the largest residual battery charge value given by $Max_{\forall k}\{Min_{i \in k}(B_i)\}$.

## 3. Simulation Parameters and Metrics

We use ns-2 (version 2.28) [18] as the simulator for our study. We implemented the FORP, LBR protocols and the MMBCR algorithm on top of DSR in ns-2. The network dimensions are 1000m x 1000m. The transmission range of each node is 250m. We vary the network density by conducting simulations with 50 nodes (low density network with an average of 10 neighbours per node) and 100 nodes (high density network with an average of 20 neighbours per node). We conduct two sets of experiments. In the first set of experiments, the energy level at each node is 1500 Joules and we ran the simulations for 1000 seconds. In the second set of experiments, the energy level at each node is 100 Joules and we ran the simulations until five node failures occur. Traffic sources are continuous bit rate (CBR). Number of source-destination (*s-d*) sessions used is 15 (low traffic load) and 30 (high traffic load). The starting times of the *s-d* sessions is uniformly distributed between 1 to 20 seconds. Data packets are 512 bytes in size; the packet sending rate is 4 data packets per second. While distributing the source-destination roles for each node, we saw to it that a node does not end up as source of more than two sessions and also not as destination for more than two sessions.

All the three routing protocols/algorithms require the use of beacon control messages to let each node advertise its presence to neighbours and learn about the neighbourhood. Beacons are exchanged for every one second. In the case of FORP, each node sends information about its location and current velocity in the beacons. Each node keeps track of the previously advertised location of its neighbour nodes. This will help to determine the direction in which the neighbour node is moving. For LBR, each node includes in the beacon packets information about the number of *s-d* sessions the node is part of. For MMBCR, each node includes in the beacon packets information about the current battery charge available at the node.

### *3.1 MAC Layer Model*

The MAC layer uses the distributed co-ordination function (DCF) of the IEEE Standard 802.11 [22] for wireless LANs. For scenarios with transmission power control, the channel negotiation is dealt as follows: the sender node transmits the Request-To-Send (RTS) packet with a transmission power corresponding to the fixed maximum transmission range of 250m. The receiver node on receiving the RTS packet, estimates the distance to the sender based on the strength of the signal received for the RTS packet. The receiver node includes this distance information in the Clear-To-Send (CTS) packet, which is transmitted with the transmission power that is just enough to reach the sender with signal strength above the receiving signal strength threshold of $3 * 10^{-10}$ W. The sender node on receiving the CTS packet uses the distance information in the CTS packet and estimates the transmission power that would be just sufficient to send the DATA packet to the receiver node. The transmission power used is calculated using the formula [19][23]: $1.1182 + 7.2 * 10^{-11}(d)^4$, *d* – distance between the transmitter and the receiver. The receiver node upon successfully receiving the DATA packet will send an ACK packet using the transmission power that was spent to send the CTS packet. The neighbours of the receiver that had earlier received the CTS packet and the neighbours of the sender that had received both the RTS and the CTS packets are free to start their own channel negotiations after they receive the ACK packet within a certain time period.

Note that the neighbour nodes of the sender that received the RTS packet and not the CTS packet within a certain time are free to start having their own transmissions while the DATA packet transmission is taking place. Similarly, the neighbours of the receiver that did not receive the CTS packet are also free to start having their own transmissions while the DATA packet transmission is taking place. Thus, transmission power control also helps us to increase the usage of bandwidth and minimize the delay in packet transmissions.

### *3.2 Energy Consumption Model*

The energy consumption at a node in an ad hoc network can be divided into three categories: (i) Energy utilized for transmitting a message, (ii) Energy utilized for receiving a message and (iii) Energy utilized in idle state. In [24], it has been shown that in the presence of overhearing, no real optimization in the energy consumption or node lifetime can be achieved. Energy consumption at a node would be dominated by the energy lost when the node is in idle state (also referred to as being in the promiscuous mode). Thus, in this paper, we do not consider the energy lost in the idle state and focus only on the energy consumed during the transmission and reception of messages (the DATA packets, the MAC layer RTS-CTS-ACK packets and the periodic beacons), and the energy consumed due to route discoveries. We model the energy consumed due to broadcast traffic and point-to-point traffic as linear functions of the packet transmission time, network density, transmission and reception powers per hop. A similar linear modelling for energy consumption has been used in [25][26]. For simulations without transmission power control, the fixed transmission power per hop is 1.4W. For simulations with transmission power control, the transmission power per hop is dynamically adjusted using the formula $1.1182 + 7.2*10^{-11}*(d)^4$, which includes power required to drive the circuit (1.1182W) and transmission power from the antenna computed using the two-ray ground reflection model [18] and distance *d* between the sender and receiver. The reception power per hop is fixed for all situations and it is 0.967W.

### *3.3 Node Mobility Model*

The node mobility model used in all of our simulations is the Random Waypoint model [20], a widely used mobility model in MANET simulation studies. According to this model, each node starts moving from an arbitrary location to a randomly selected destination location at a speed uniformly distributed in the range $[v_{min},…,v_{max}]$. Once the destination is reached, the node may stop there for a certain time called the pause time and then continue to move by choosing a different target location and a different velocity. In this paper, we set $v_{min} = 0$, and each node chooses speed uniformly distributed between 0 and $v_{max}$. The $v_{max}$ values used are 5, 10 and 20 m/s (representing low node mobility scenarios) and 30, 40 and 50m/s (representing high node mobility scenarios). Pause time is 0 seconds.

### *3.4 Performance Metrics*

Each data point in Figures 1 through 16 is an average of data collected using 5 mobility trace files and 5 sets of randomly selected 15 and 30 *s-d* sessions. We study the following performance metrics for the three routing protocols:

(i) *Number of route transitions* − the average of the number of route discoveries per *s-d* session, averaged over all the *s-d* sessions of a simulation run.
(ii) *Hop count per route* − average of the number of hops in the routes of an *s-d* session, time-averaged after considering all the *s-d* sessions. For example, if a routing protocol uses paths P1 and P2 of hop counts 3 and 5 for time 10 seconds and 5 seconds respectively, then the time-averaged hop count for the total time of 15 seconds is (3*10 + 5*5)/15 = 3.67 seconds and not simply 4 seconds.
(iii) *End-to-end delay per data packet* − average of the delay incurred by the data packets that originate at the source and delivered at the destination. The delay incurred by a data packet includes all the possible delays − the buffering delay due to the route acquisition latency, the queuing delay at the interface queue to access the medium, transmission delay, propagation delay, and the retransmission delays due to the MAC layer collisions.
(iv) *Energy consumed per node* − average of the energy consumed across all the nodes in the network. The energy consumed at a node includes the energy lost due to the transmission and reception of data packets, MAC layer packets, periodic beacons exchange, and the energy consumed due to route discoveries.
(v) *Fairness of node usage* − measured using the standard deviation of the energy consumed per node, which is the square root of the average of the squares of the difference between the energy consumed at each node and the averaged energy consumed per node. Ideally, the value of this metric should be zero to indicate that all nodes have been used fairly and no node is overused.
(vi) *Time of first node failure* − The time of first node failure due to the exhaustion of battery charge during the simulation with a particular routing protocol.
(vii) *Time since the first node failure* − We measure the failure times for the first 5 node failures. For the $2^{nd}$, $3^{rd}$, $4^{th}$ and the $5^{th}$ node failures, we measure the relative time these failures happened since the time of first node failure. This helps us to study whether a single node failure simultaneously triggers a sequence of node failures or whether there is sufficient time between successive node failures.

## 4. Simulation Results

In this section, we present the simulation results obtained for each of the above seven performance metrics with respect to the three routing protocols studied in this paper.

### *4.1 Route Transitions*

In all the simulation conditions, FORP incurs the least number of route transitions, followed by LBR and MMBCR, in this order. For a given network density and node mobility, there is no appreciable change in the number of route transitions, as we increase the offered data traffic load from low to high. This is attributed to the fact that the *s-d* sessions are independent of each other. As we increase the maximum node velocity from 5 m/s to 50 m/s, the number of route transitions for each of the three protocols increases by a factor of 10 for low-density networks (refer Figures 1.1, 2.1, 3.1 and 4.1) and by a factor of 7 to 8 for high-density networks (refer Figures 5.1, 6.1, 7.1 and 8.1). We find that there is no significant change in the number of route transitions incurred by the three routing protocols when operated with and without power control (refer Figures 1.1, 2.1, 3.1, 4.1, 5.1, 6.1, 7.1 and 8.1). For a given simulation condition, the number of route transitions incurred by MMBCR is 15 to 25% more than that of LBR.

The number of route transitions incurred by FORP is the minimum and this is as expected because it is a stable path routing protocol and chooses the route that has the largest predicted lifetime since the time of selection. The number of route transitions incurred by FORP in high-density networks is often greater than that incurred in low-density networks by a factor of 5 to 25%. In high-density networks, as the number of nodes within the neighbourhood is increased, FORP gets more chances of finding stable links with longer predicted lifetime.

In networks of low density, the number of route transitions incurred by LBR is about 3 to 4 times to that of FORP under networks of low mobility and high mobility. In networks of high density, the number of route transitions incurred by LBR is about 5 to 7 times to that of FORP under low node mobility and is about 4 to 5 times to that of FORP under high node mobility. In networks of low density, the number of route transitions incurred by MMBCR is about 4 to 5 times to that of FORP under low node mobility and is about 5 to 5.5 times to that of FORP under high node mobility. In networks of high density, the number of route transitions incurred by MMBCR is about 6 to 7.5 times to that of FORP under low node mobility and 5 to 6.5 times to that of FORP under high node mobility.

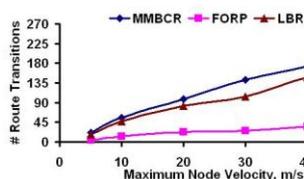
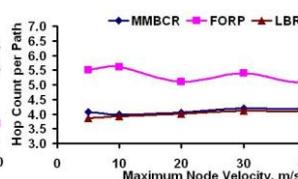
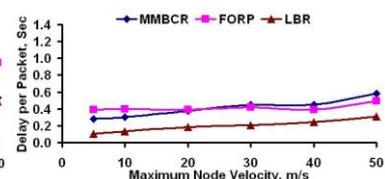

**Fig 1.1:** # Route Transitions    **Fig 1.2:** Hop Count per Path    **Fig 1.3:** Delay per Packet

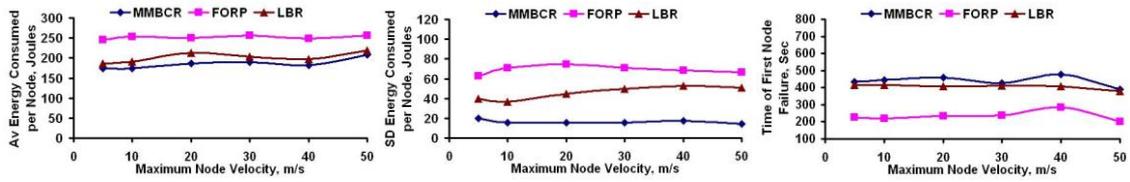

**Fig 1.4:** Energy Cons. per Node   **Fig 1.5:** Node Fairness   **Fig 1.6:** Time of 1st Node Failure

**Figure 1:** Performance Metrics (50 nodes, 15 *s-d* pairs, Absence of Power Control)

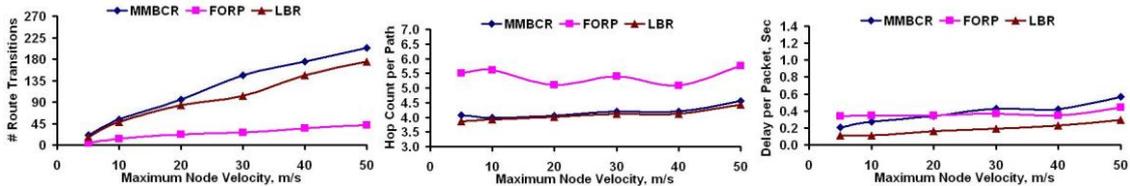

**Fig 2.1:** # Route Transitions   **Fig 2.2:** Hop Count per Path   **Fig 2.3:** Delay per Packet

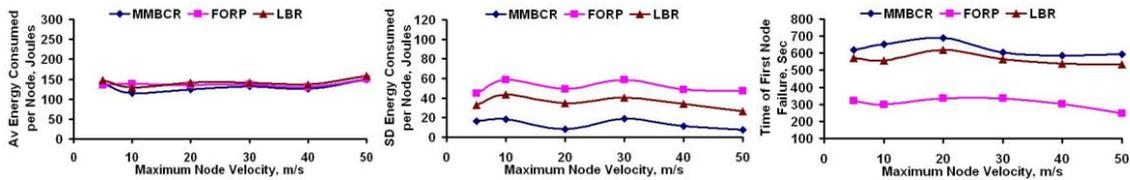

**Fig 2.4:** Energy Cons. per Node   **Fig 2.5:** Node Fairness   **Fig 2.6:** Time of 1st Node Failure

**Figure 2:** Performance Metrics (50 nodes, 15 *s-d* pairs, Presence of Power Control)

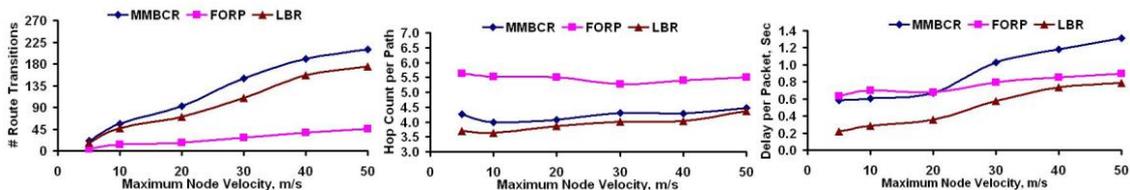

**Fig 3.1:** # Route Transitions   **Fig 3.2:** Hop Count per Path   **Fig 3.3:** Delay per Packet

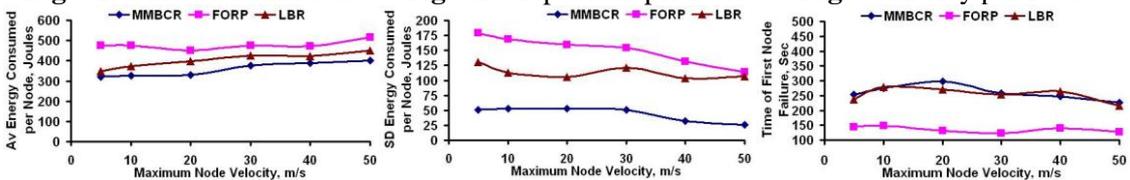

**Fig 3.4:** Energy Cons. per Node   **Fig 3.5:** Node Fairness   **Fig 3.6:** Time of 1st Node Failure

**Figure 3:** Performance Metrics (50 nodes, 30 *s-d* pairs, Absence of Power Control)

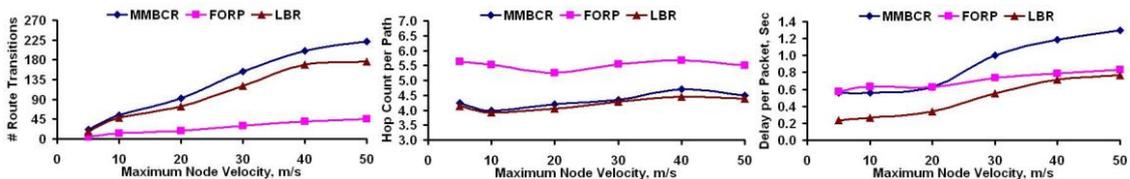

**Fig 4.1:** # Route Transitions   **Fig 4.2:** Hop Count per Path   **Fig 4.3:** Delay per Packet

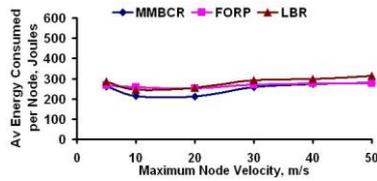 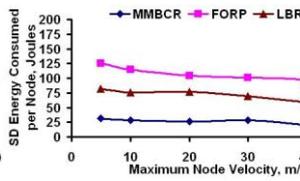 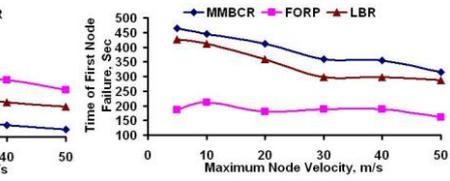

**Fig 4.4:** Energy Cons. per Node      **Fig 4.5:** Node Fairness      **Fig 4.6:** Time of 1st Node Failure

**Figure 4:** Performance Metrics (50 nodes, 30 *s-d* pairs, Presence of Power Control)

*4.2 Hop Count per Path*

For all the simulation conditions (refer Figures 1.2, 2.2, 3.2, 4.2, 5.2, 6.2, 7.2 and 8.2), we observe LBR to incur the minimum number of hops, closely followed by MMBCR. LBR tries to minimize the sum of traffic interferences of the constituent intermediate nodes of a route and this helps to minimize the number of intermediate nodes that form the route. For the source and destination of a route, the traffic interference value is considered to be zero as these nodes have to transmit and receive the data packets at any cost. MMBCR also attempts to maximize the bottleneck battery charge of a route. It attempts to avoid intermediate nodes that will have a lower bottleneck battery charge. Note that in MMBCR, the battery charge available at the source and destination nodes is not considered while choosing the bottleneck path battery charge. Hence, trying to choose a path with the largest bottleneck battery charge helps indirectly to find a path that has minimum hop count. For a given network density and offered data traffic load, the hop count of the MMBCR paths under high node mobility conditions is about 20% more than that incurred under low node mobility conditions. This also contributes to the relative increase in the number of route transitions incurred by MMBCR at high node mobility.

FORP incurs the maximum number of hops for all the simulation conditions. FORP in its pursuit to find stable paths with the maximum predicted route lifetime, attempts to choose links that have a larger predicted lifetime. As the distance separating the constituent nodes of a link is reduced, the probability of the link to have a larger predicted lifetime increases. Links in which its constituent nodes are moving close to each other are predicted to exist for a longer time than links in which the constituent nodes are moving away from each other or moving parallel to each other. As FORP prefers to connect the source and destination nodes using links with shorter physical distance (because such links have a higher predicted lifetime), the hop count of FORP paths is higher. It is also interesting to observe from the figures that the hop count of FORP paths in high-density networks is greater than that incurred for low-density networks by a factor of 15 to 25%. This is because, as we increase the network density, the number of nodes in a neighbourhood increases and FORP gets a larger pool of links to choose from. In networks of higher density, FORP manages to find more stable links, but the hop count increases accordingly. The hop count of FORP paths is 30 to 40% and 60 to 70% of the hop count of LBR paths in networks of low density and high density respectively.

Similar to the case of route transitions, we observe that the hop count per path is not affected by power control (compare Figures 1.2, 2.2, 3.2, 4.2, 5.2, 6.2, 7.2 and 8.2). Also, for a fixed network density, the hop count per path is not much affected by the offered data traffic load as the *s-d* sessions are independent of each other (compare Figure 1.2 with Figure 3.2, Figure 5.2 with 7.2, Figure 2.2 with Figure 4.2 and Figure 6.2 with Figure 8.2). The hop count per path for LBR is slightly lower in high-density networks compared to that incurred in low-density networks. This is attributed to the larger pool of intermediate nodes available to choose from so that the sum of the traffic interferences can be minimized. As the number of nodes in the neighbourhood increases, few intermediate nodes can be chosen that would be sufficient enough to connect the source and destination such that the hop count is minimized. The above results indicate a clear tradeoff between route stability and hop count. FORP incurs the minimum number of route transitions and MMBCR incurs the maximum number of route transitions, closely followed by LBR. On the other hand, LBR and MMBCR incur lower hop count compared to that of FORP.

### *4.3 End-to-End Delay per Data Packet*

Figures 1.3, 2.3, 3.3, 4.3, 5.3, 6.3, 7.3 and 8.3 indicate that the end-to-end delay per data packet for LBR is the lowest among the three protocols simulated. This is attributed to the fact that LBR attempts to minimize the sum of the traffic interferences of the nodes in a path. LBR paths have the minimum number of intermediate nodes as seen in the results for hop count. As a result, LBR attempts to route the packets through nodes that are least congested and through the minimum number of intermediate nodes. With respect to the other two protocols, we observe that the end-to-end delay per data packet for MMBCR is smaller than that of FORP under low node mobility conditions and found to be usually larger than that of FORP under high node mobility conditions. We attribute this to the significant increase in the number of MMBCR route transitions and also an appreciable increase in the hop count per path under high node mobility conditions. The end-to-end delay per data packet is the sum of the delays incurred for hop-to-hop data packet transfer, link-layer access delays, route acquisition delays and queuing delays averaged over all the data packets transferred from the source to the destination.

We see that the end-to-end delay per data packet incurred by the routing protocols is slightly influenced by power control. FORP is the most influenced among the three. This is because FORP paths have a larger hop count, but the physical length of each hop is smaller. As a result, power control helps to avoid the forwarding and receiving interference of the data traffic load at nodes that are outside the radius of the hop length but within the transmission range of the nodes. This effect is felt at all the constituent hops of FORP route. Since there are more hops in FORP routes, the difference in the end-to-end delay per data packet in the presence and absence of power control is noticeable. For a given offered data traffic load, node mobility and network density, the end-to-end delay per data packet in the absence of power control is about 10-15% than that incurred in the presence of power control. In the case of LBR, the influence of power control is observed in networks of high density. The reduction in the end-to-end delay per data packet for LBR and MMBCR in the presence of control is by 10-15% at low node mobility and below 10% at high node mobility. The reduction in the queuing delay per data packet at high node mobility for LBR and MMBCR is offset by the increase in the route-acquisition delay and the frequent route discoveries.

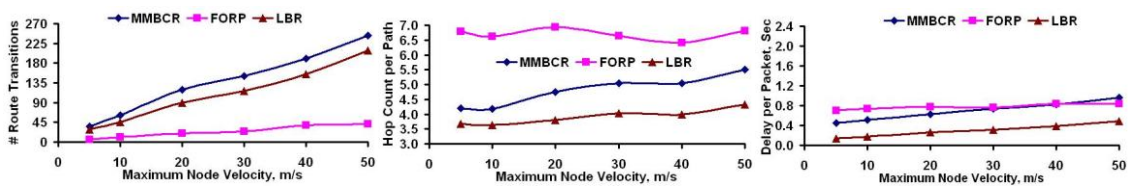

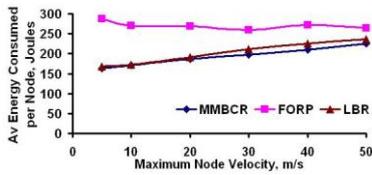
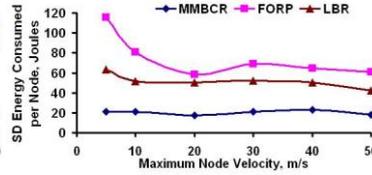
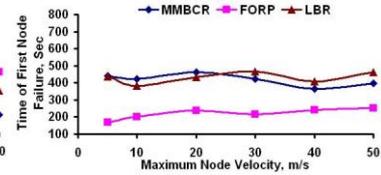

**Fig 5.1:** # Route Transitions    **Fig 5.2:** Hop Count per Path    **Fig 5.3:** Delay per Packet

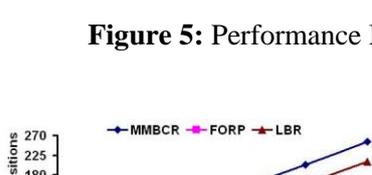
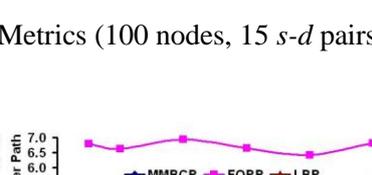
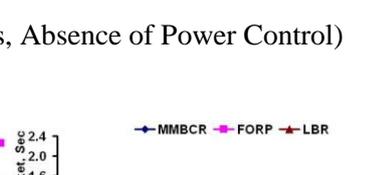

**Fig 5.4:** Energy Cons. per Node    **Fig 5.5:** Node Fairness    **Fig 5.6:** Time of 1st Node Failure

**Figure 5:** Performance Metrics (100 nodes, 15 *s-d* pairs, Absence of Power Control)

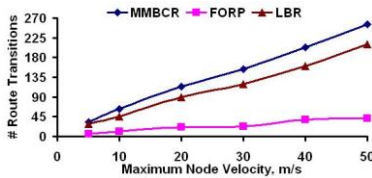
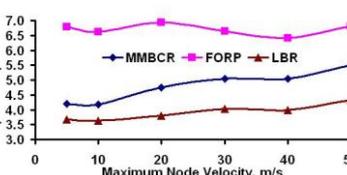
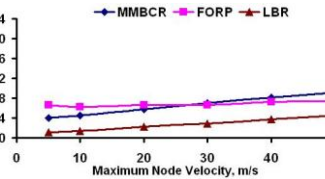

**Fig 6.1:** # Route Transitions    **Fig 6.2:** Hop Count per Path    **Fig 6.3:** Delay per Packet

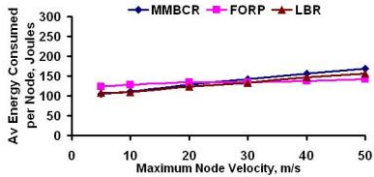
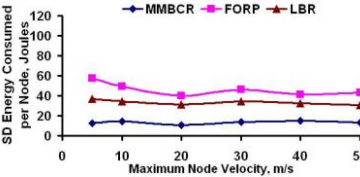
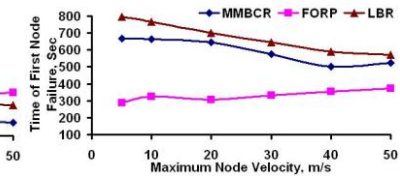

**Fig 6.4:** Energy Cons. per Node    **Fig 6.5:** Node Fairness    **Fig 6.6:** Time of 1st Node Failure

**Figure 6:** Performance Metrics (100 nodes, 15 *s-d* pairs, Presence of Power Control)

### *4.4 Energy Consumed per Node*

The energy consumed per node is the sum of the energy lost at a node due to data packet transmission and reception, channel negotiation at the link layer, periodic beacon exchange, broadcast route discovery packet transmission and reception and transmission and reception of other control packets, across all the source-destination sessions, over all of the simulation runs. Figures 1.4, 2.4, 3.4, 4.4, 5.4, 6.4, 7.4 and 8.4 respectively illustrate the energy consumed per node for the three routing protocols in the absence and presence of power control. In the absence of power control, the energy consumed per node is the least for MMBCR, closely followed by LBR. This could be attributed to the nature of MMBCR to choose the route with the maximum value for the bottleneck battery charge and not to unnecessarily burden nodes with minimum battery charge. Also, the tendency of MMBCR to directly establish the path from the source to destination, if they are directly reachable, helps to avoid involving other nodes unnecessarily as intermediate nodes. FORP incurs the maximum energy consumption per node in the absence of power control. This is attributed to the presence of relatively more hops connecting the source and the destination (i.e., more intermediate nodes forwarding the data packets for others).

In the absence of power control, the energy consumed per node is directly proportional to the number of hops traversed by the data packet. For a given network density, for each of the three routing protocols, the energy consumed per node is almost doubled as the offered data traffic load is doubled from 15 *s-d* pairs to 30 *s-d* pairs. In low-density networks and for a given offered data packet traffic load, the energy consumed per node for FORP and LBR is respectively 1.2 to 1.4 and 1.05 to 1.2 times of the energy consumed per node for MMBCR. In high-density networks and for a given offered data traffic load, the energy consumed per node for FORP is 1.4 to 1.7 and 1.2 to 1.4 times to that of MMBCR under conditions of low node mobility and high node mobility respectively. On the other hand, in high-density networks and for a given offered traffic load, the energy consumed per node for LBR is at most 1.15 times to that of MMBCR. The increase in the energy usage per node for FORP at high node density is attributed to the usage of longer hop paths to maximize the route lifetime. For a given offered data traffic load, the energy consumed per data packet for each of the three routing protocols is almost the same as the network density is doubled. For a given offered data traffic load, the energy consumed per data packet for each of MMBCR and LBR at maximum node velocity of 5m/s is about 20% and 30 to 40% more than that incurred by these two routing protocols at maximum node velocity of 50 m/s in networks of low and high density respectively. The energy consumed per data packet for FORP is almost the same for both low and high node mobility.

In the presence of power control, the energy consumed per node for FORP is tremendously reduced and is only about 48% of that incurred in the absence of power control. On the other hand, for MMBCR and LBR, the energy consumed per node in the presence of power control is about 70% of that incurred in the absence of power control. The energy consumed per node is the least reduced for LBR and MMBCR because of the physical length of its hops being close to the transmission range of the nodes. On the other hand, for FORP, the physical length of the hops is about 50 to 60% of the transmission range of the nodes. Hence, we find relatively higher effectiveness while using power control for FORP.

For a given offered data traffic load, in networks of low density, the increase in the energy consumed per node for all the three routing protocols in the presence of power control is at most by a factor of only 10% as the maximum node velocity is increased from 5 m/s to 50 m/s. On the other hand, for a given offered traffic load, in high-density networks, the increase in the energy consumed per node for MMBCR and LBR in the presence of power control is by a factor of 60% and 45% respectively as the maximum node velocity is increased from 5 m/s to 50 m/s. The increase in energy consumed per node in the presence of power control for FORP in high density networks is 7 to 15% as the maximum node velocity is increased from 5 m/s to 50 m/s. Thus, FORP is the most scalable routing protocol, both in the absence and presence of power control, with respect to energy consumed per node as node mobility is increased.

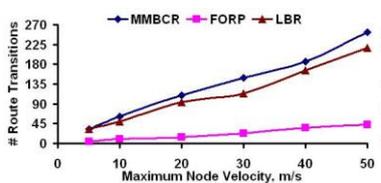 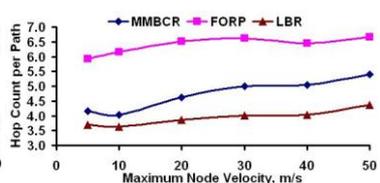 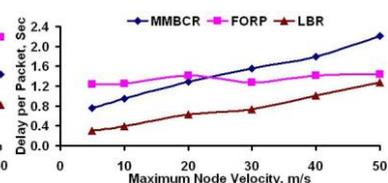

**Fig 7.1:** # Route Transitions     **Fig 7.2:** Hop Count per Path     **Fig 7.3:** Delay per Packet

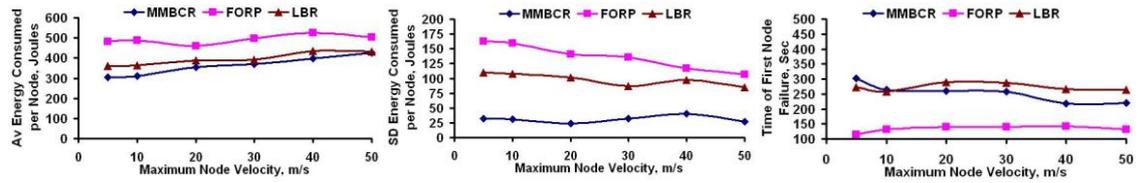

**Fig 7.4:** Energy Cons. per Node    **Fig 7.5:** Node Fairness    **Fig 7.6:** Time of 1st Node Failure

**Figure 7:** Performance Metrics (100 nodes, 30 *s-d* pairs, Absence of Power Control)

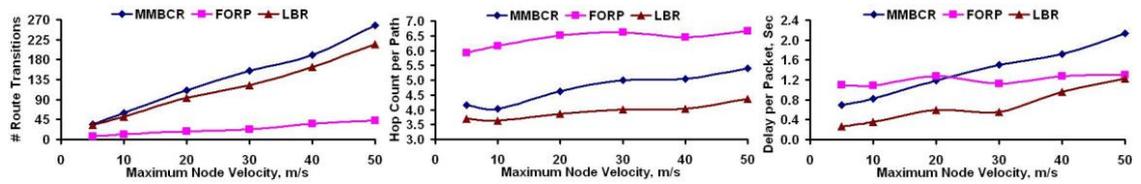

**Fig 8.1:** # Route Transitions    **Fig 8.2:** Hop Count per Path    **Fig 8.3:** Delay per Packet

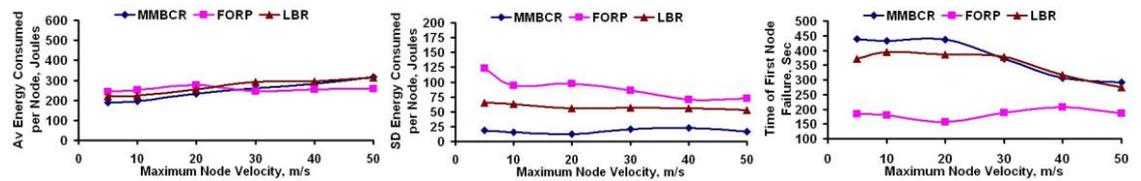

**Fig 8.4:** Energy Cons. per Node    **Fig 8.5:** Node Fairness    **Fig 8.6:** Time of 1st Node Failure

**Figure 8:** Performance Metrics (100 nodes, 30 *s-d* pairs, Presence of Power Control)

## *4.5 Fairness of Node Usage*

The fairness of node usage for the three routing protocols is measured by computing the standard deviation of the energy consumed per node. Ideally, to say that all the nodes have been fairly used by the routing protocol, we would expect the standard deviation of the energy consumed per node to be zero. But due to the stochastic nature of ad hoc networks, not all nodes are being fairly used. Certain nodes are observed to be preferentially used than other nodes. Figures 1.5, 2.5, 3.5, 4.5, 5.5, 6.5, 7.5 and 8.5 respectively illustrate the standard deviation of energy consumed per node for the three routing protocols in the absence and presence of power control. The smaller is the value of the standard deviation of node usage, the better the fairness of node usage of the routing protocol.

Figures 1.5, 2.5, 3.5, 4.5, 5.5, 6.5, 7.5 and 8.5 illustrate that MMBCR is the most fair among the three routing protocols. FORP is the most unfair of all the three and LBR is in between. The fairness of MMBCR is justified by the fact that it attempts to divert routes from heavily used nodes (in terms of energy consumed) towards lightly used nodes. MMBCR chooses routes such that the bottleneck battery charge of the route is the maximum. So, MMBCR cleverly avoids from over utilizing nodes, when there are nodes that are under utilized. FORP incurs the highest standard deviation of energy usage because, stable paths tend to exist for a long time and use certain set of nodes preferentially over other nodes.

For all the three routing protocols, the standard deviation of energy consumed per node decreases because of power control. This is attributed to the relatively lower energy consumed at all the nodes because of power control when compared with the energy consumed in the absence of power control. This in directly indicates that the fairness of node usage of the routing protocols improves with power control. For a low offered data traffic load, the standard deviation of energy consumed for each of the three routing protocols in the absence of power control is about 1.35 (low-density) and 1.55 (high-density) times to that observed in the presence of power control. For a high offered data traffic load and a given network density, the standard deviation of energy consumed per node for MMBCR, FORP and LBR in the absence of power control is respectively 1.8, 1.5, 1.7 times to that observed in the presence of power control. With respect to the fairness of node usage, MMBCR gets the maximum benefit from using power control and FORP gets a relatively lower benefit from using power control.

We also observe that the standard deviation of energy consumed per node for all the three routing decreases with increase in node mobility. This could be attributed to increase in the number of route failures as we increase node mobility. In the absence of power control, for a given network density and offered data traffic load, the network lifetime of MMBCR, FORP and LBR at the maximum node velocity values of 50 m/s is respectively about 72%, 72% and 88% of that incurred for the three routing protocols at the maximum node velocity values of 5 m/s. In the presence of power control, for a given network density and offered data traffic load, the network lifetime of MMBCR, FORP and LBR at the maximum node velocity values of 50 m/s is respectively about 70%, 76% and 77% of that incurred for the three routing protocols at the maximum node velocity values of 5 m/s.

*4.6 Time of First Node Failure*

The fairness of node usage for the three routing protocols is measured by computing the standard deviation of the energy consumed per node. Ideally, to say that all the nodes have been fairly used by the routing protocol, we would expect the standard deviation of the energy consumed per node to be zero. But due to the stochastic nature of ad hoc networks, not all nodes are being fairly used. Certain nodes are observed to be preferentially used than other nodes. Figures 1.5, 2.5, 3.5, 4.5, 5.5, 6.5, 7.5 and 8.5 respectively illustrate the standard deviation of energy consumed per node for the three routing protocols in the absence and presence of power control. The smaller is the value of the standard deviation of node usage, the better the fairness of node usage of the routing protocol.

Performance data for time of node failures are collected when the available battery charge at each node is 100 Joules. Figures 1.6, 2.6, 3.6, 4.6, 5.6, 6.6, 7.6 and 8.6 illustrate that the time of first node failure is high for both LBR and MMBCR and close enough to each other for most of the simulation conditions. This could be due to the design of these routing protocols to prefer nodes that have been under utilized over nodes that have been over utilized. The time of first node failure is low for FORP in all the cases. This could be attributed to the preferential usage of nodes lying on the stable path and a larger hop count per path. For a given offered traffic load and network density, the time of first node failure for FORP is about 57% and 47% of the time of first node failure for MMBCR in the absence and presence of power control respectively.

All the three protocols exhibit an improvement in the time of 1st node failure while using power control. For a given network density and offered data traffic load, the time of first node failure for all the three routing protocols in the presence of power control is roughly about, on average, 1.4 times to that incurred in the absence of power control. For all the three protocols, for a given offered traffic load, there is no significant improvement in the time of first node failure, as we double the network density, both in the presence and in the absence of power control.

We observe that in the absence of power control, for a given network density and offered data traffic load, the time of first node failure for MMBCR, FORP and LBR at the maximum node velocity values of 50 m/s is respectively about 85%, 110% and 95% of that incurred at the maximum node velocity values of 5 m/s. On the other hand, in the presence of power control, for a given network density and offered data traffic load, the time of first node failure for MMBCR, FORP and LBR at the maximum node velocity values of 50 m/s is respectively about 77%, 98% and 77% of that incurred at the maximum node velocity values of 5 m/s. This illustrates that the MMBCR and LBR protocols make maximum use of power control by conserving the battery charge at the nodes while transferring data packets and use that energy for route discovery at high node mobility. FORP is not much influenced by the use of power control at high node mobility. This is because, the increase in the number of route transitions for FORP is the lowest with increase in node mobility and the routing protocol spends most of the energy in transferring the data packets through stable paths of larger hop count.

In the absence of power control, for a given network density, the time of first node failure incurred with low offered data traffic load for the three routing protocols is about 1.5 (low node mobility) to 1.8 (high node mobility) times to that incurred for high offered data traffic load. In the presence of power control, for a given network density, the time of first node failure incurred with low data traffic load for the three routing protocols is about 1.35 (low node mobility) to 1.9 (high node mobility) times to that incurred for high offered data traffic load.

### *4.7 Time since the First Node Failure*

Figures 9 through 16 illustrate the time of node failures since the first node failure in the presence and absence of power control for the different conditions of offered data traffic load, network densities and node mobility. We measure up to the time of the 5th node failure. Note that the time values in Figures 9 through 16 are not the absolute failure times. The time values in these figures illustrate how long beyond the first node failure, that the second node fails, the third node fails, and etc. Hence, these times are relative to the time of first node failure for each routing protocol, that is, the time of first node failure in all of Figures 9 through 16 is 0. The actual time of first node failure is illustrated in Figures 1.6, 2.6, 3.6, 4.6, 5.6, 6.6, 7.6 and 8.6. We make the following significant observations from Figures 9 through 16:

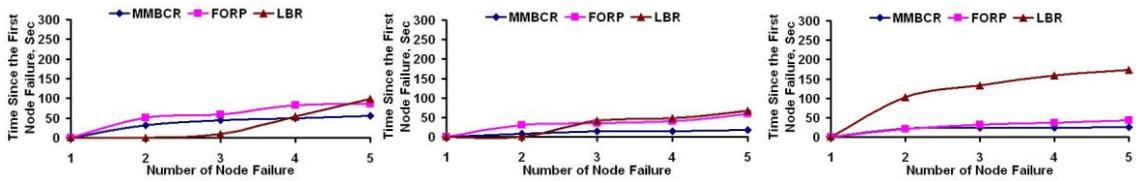

**Fig 9.1**: $v_{max}$ = 5 m/s    **Fig 9.2**: $v_{max}$ = 10 m/s    **Fig 9.3**: $v_{max}$ = 20 m/s

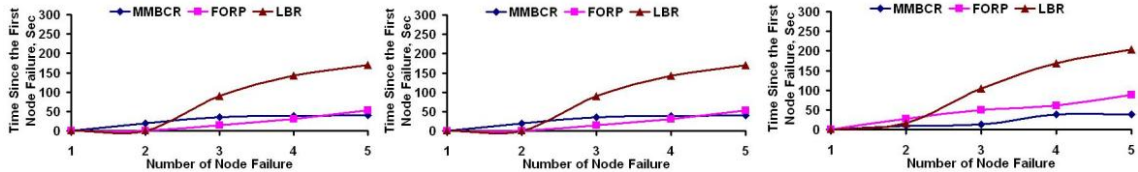

**Fig 9.4**: $v_{max}$ = 30 m/s    **Fig 9.5**: $v_{max}$ = 40 m/s    **Fig 9.6**: $v_{max}$ = 50 m/s

**Figure 9:** Time since First Node Failure (50 Nodes, 15 *s-d* Pairs, Absence of Power Control)

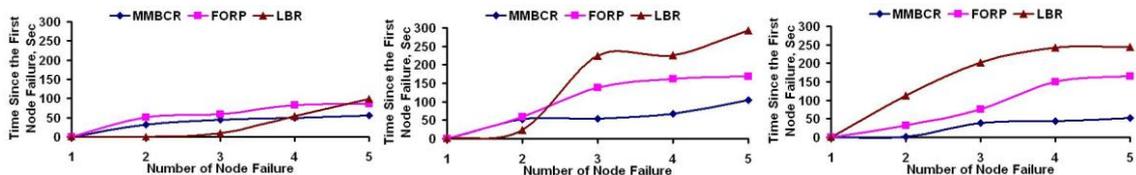

**Fig 10.1**: $v_{max}$ = 5 m/s    **Fig 10.2**: $v_{max}$ = 10 m/s    **Fig 10.3**: $v_{max}$ = 20 m/s

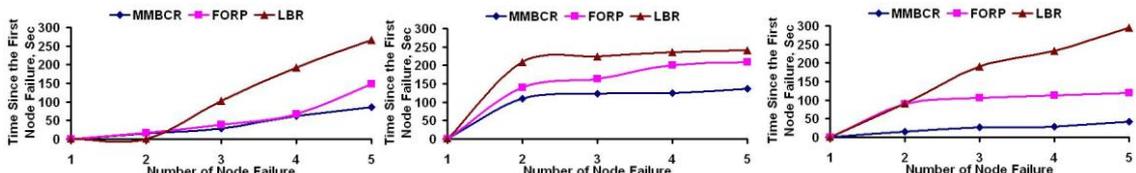

**Fig 10.4**: $v_{max}$ = 30 m/s    **Fig 10.5**: $v_{max}$ = 40 m/s    **Fig 10.6**: $v_{max}$ = 50 m/s

**Figure 10:** Time since First Node Failure (50 Nodes, 15 *s-d* Pairs, Presence of Power Control)

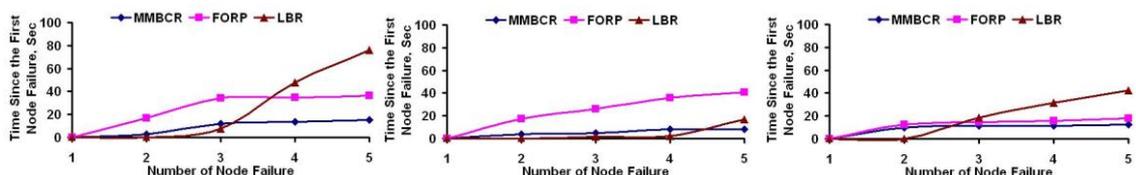

**Fig 11.1**: $v_{max}$ = 5 m/s    **Fig 11.2**: $v_{max}$ = 10 m/s    **Fig 11.3**: $v_{max}$ = 20 m/s

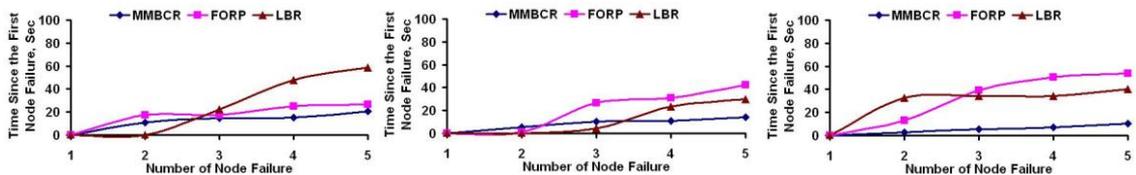

**Fig 11.4**: $v_{max}$ = 30 m/s    **Fig 11.5**: $v_{max}$ = 40 m/s    **Fig 11.6**: $v_{max}$ = 50 m/s

**Figure 11:** Time since First Node Failure (50 Nodes, 30 *s-d* Pairs, Absence of Power Control)

The lifetime of the nodes, after the 1st node failure, is least improved for MMBCR in almost all the simulation conditions. This could be attributed to the fact that MMBCR attempts to equally exploit the battery charge at all the nodes in the network. From Figures 1.5, 2.5, 3.5, 4.5, 5.5, 6.5, 7.5 and 8.5, we observe MMBCR is the most fair among the three routing protocols. So, even though the time of 1st node failure for MMBCR is the highest of the three routing protocols, the battery charge at the other nodes has also gone down at the time of first node failure, leading to subsequent node failures within a short time.

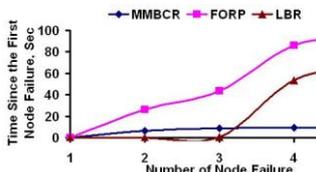 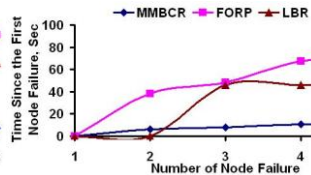 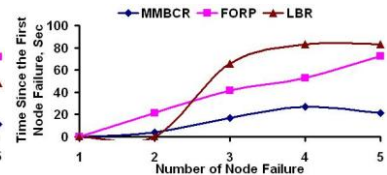

**Fig 12.1**: $v_{max}$ = 5 m/s    **Fig 12.2**: $v_{max}$ = 10 m/s    **Fig 12.3**: $v_{max}$ = 20 m/s

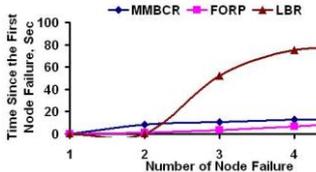 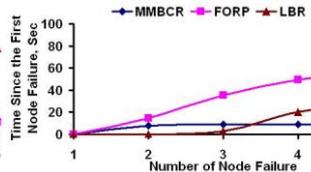 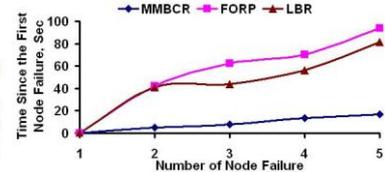

**Fig 12.4**: $v_{max}$ = 30 m/s    **Fig 12.5**: $v_{max}$ = 40 m/s    **Fig 12.6**: $v_{max}$ = 50 m/s

**Figure 12:** Time since First Node Failure (50 Nodes, 30 *s-d* Pairs, Presence of Power Control)

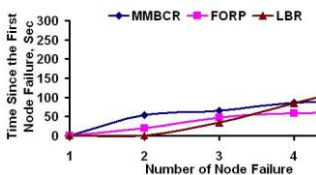 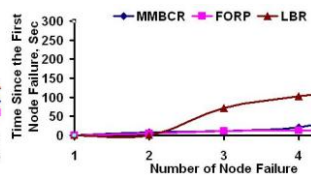 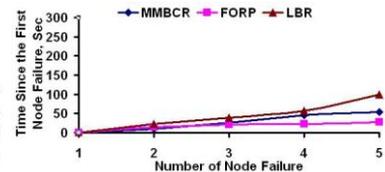

**Fig 13.1**: $v_{max}$ = 5 m/s    **Fig 13.2**: $v_{max}$ = 10 m/s    **Fig 13.3**: $v_{max}$ = 20 m/s

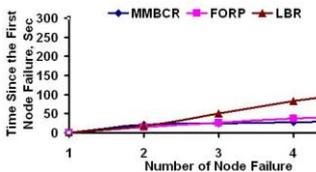 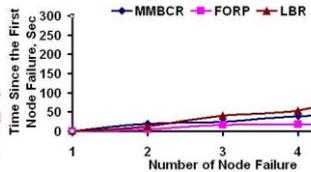 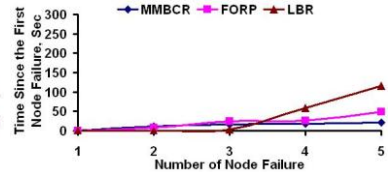

**Fig 13.4**: $v_{max}$ = 30 m/s    **Fig 13.5**: $v_{max}$ = 40 m/s    **Fig 13.6**: $v_{max}$ = 50 m/s

**Figure 13:** Time since First Node Failure (100 Nodes, 15 *s-d* Pairs, Absence of Power Control)

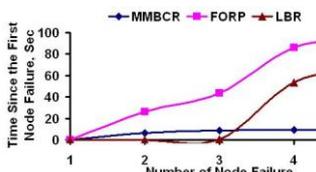 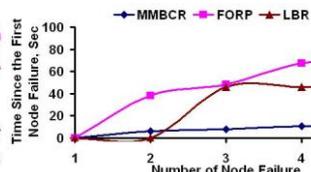 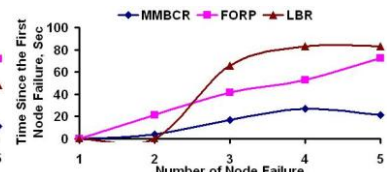

**Fig 14.1**: $v_{max}$ = 5 m/s    **Fig 14.2**: $v_{max}$ = 10 m/s    **Fig 14.3**: $v_{max}$ = 20 m/s

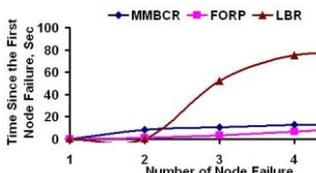 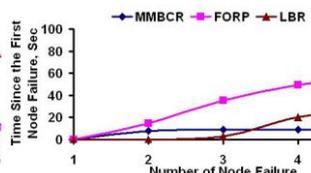 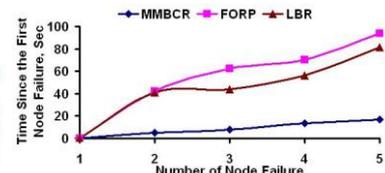

**Fig 14.4**: $v_{max}$ = 30 m/s    **Fig 14.5**: $v_{max}$ = 40 m/s    **Fig 14.6**: $v_{max}$ = 50 m/s

**Figure 14:** Time since First Node Failure (100 Nodes, 15 *s-d* Pairs, Presence of Power Control)

For scenarios of low offered data traffic load and low density, the time of 5$^{th}$ node failure is extended for a maximum of 300 seconds (at high node mobility) beyond the time of first node failure in the presence of power control. In the absence of power control, the time of 5$^{th}$ node failure is extended for a maximum of 200 seconds (at high node mobility). The maximum extensions are possible at high node mobility, as mobility naturally helps to balance the forwarding load at the nodes. On the top of this, the load balancing characteristic of LBR helps the routing protocol to extend the lifetime of the nodes to the maximum.

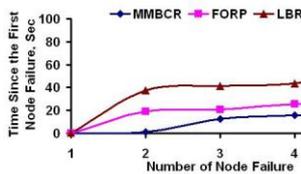 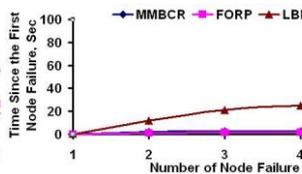 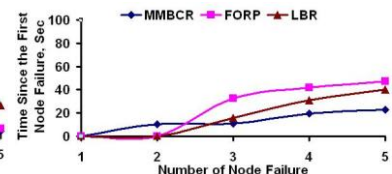
**Fig 15.1**: $v_{max} = 5$ m/s    **Fig 15.2**: $v_{max} = 10$ m/s    **Fig 15.3**: $v_{max} = 20$ m/s

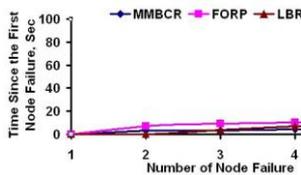 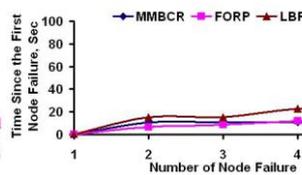 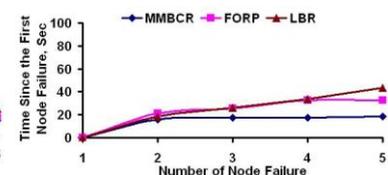
**Fig 15.4**: $v_{max} = 30$ m/s    **Fig 15.5**: $v_{max} = 40$ m/s    **Fig 15.6**: $v_{max} = 50$ m/s

**Figure 15:** Time since First Node Failure (100 Nodes, 30 *s-d* Pairs, Absence of Power Control)

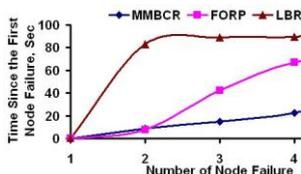 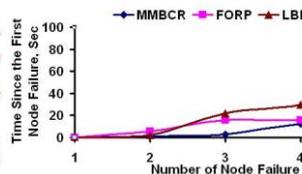 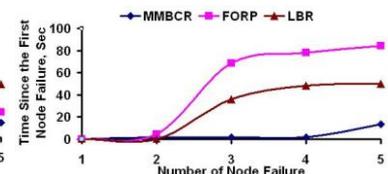
**Fig 16.1**: $v_{max} = 5$ m/s    **Fig 16.2**: $v_{max} = 10$ m/s    **Fig 16.3**: $v_{max} = 20$ m/s

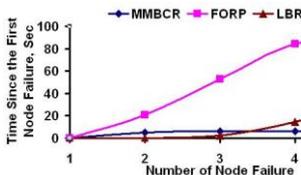 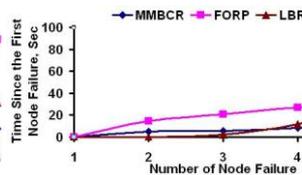 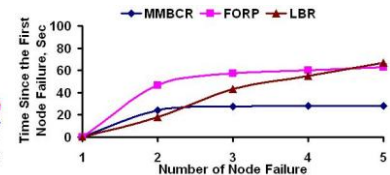
**Fig 16.4**: $v_{max} = 30$ m/s    **Fig 16.5**: $v_{max} = 40$ m/s    **Fig 16.6**: $v_{max} = 50$ m/s

**Figure 16:** Time since First Node Failure (100 Nodes, 30 *s-d* Pairs, Presence of Power Control)

For scenarios of high offered data traffic load, for both low and high low density, the time of 5$^{th}$ node failure is extended for a maximum of 100 seconds (at low node mobility) beyond the time of first node failure in the presence of power control. In the absence of power control, the time of 5$^{th}$ node failure is extended for a maximum of 65 to 75 seconds (at low node mobility). When the offered data traffic load is high, the maximum extension in the failure times of the nodes is possible at low mobility, compared to the high mobility scenarios. This is attributed to the increased energy consumption, especially in networks of high density, due to route discoveries for several source-destination sessions (30, high data traffic load) at high node mobility.

For scenarios of low offered traffic load and high density, the time of $5^{th}$ node failure is extended for a maximum of 170 seconds beyond the time of first node failure in the presence of power control and to a maximum time of 135 seconds beyond the time of first node failure in the absence of power control. These extensions are mostly possible in networks of low mobility, because route discovery through flooding in high density networks is energy-expensive.

Though FORP incurs the lowest time of first node failure, the time of $5^{th}$ node failure is significantly beyond the time of first node failure, especially in the presence of power control. FORP preferentially uses nodes that lie on the stable path. When the node(s) in this path die due to exhaustion of battery charge, FORP resorts to using other nodes in the network and due to the unfairness in node usage, chances are that nodes that have not been used heavily begin to be used after certain heavily used nodes fail.

**5. Summary of Simulation Results and Conclusions**

The high-level contribution of this paper is a simulation-based performance comparison analysis of three different categories of MANET routing protocols: Stability-based FORP, Power-aware MMBCR and the Load-balancing routing (LBR) protocol. The simulations have been conducted under different scenarios of node density, node mobility, offered data traffic load and in the presence/absence of power control. Some of the key conclusions on the performance results are summarized below:

Tradeoff between stability and hop count: FORP incurs the least number of route transitions among the three routing protocols for all simulation conditions. MMBCR incurs the maximum number of route transitions, closely followed by LBR. LBR incurs the minimum number of hops per path, closely followed by MMBCR. FORP incurs the maximum number of hops per path for all the simulation conditions. FORP routes are more stable in networks of higher density compared to networks of lower density. The tradeoff is the increase in hop count in networks of high density compared to those in low density. For a given network density, there is no significant in the number of route transitions and hop count per path for each of the three routing protocols when operated with and without power control and with increase in the offered data traffic load from 15 *s-d* pairs to 30 *s-d* pairs.

LBR incurs the lowest end-to-end delay per data packet among the three routing protocols for all the simulation conditions tested. The end-to-end delay per data packet for MMBCR is smaller than that of FORP under low node mobility conditions and larger than that of FORP under high node mobility conditions. The end-to-end delay per data packet incurred by FORP is the most influenced by power control. When power control is conducted on hops with smaller physical length, we manage to reduce the forwarding and receiving interference of the data traffic load at the non-participating nodes to a maximum. For a given network density and offered data traffic load, as we increase the node mobility, FORP has the slowest increase in the end-to-end delay per data packet. For low and high network density, FORP and LBR are respectively the most scalable with respect to end-to-end delay per data packet as we increase the offered data traffic load. For low offered data traffic load and high offered data traffic load, LBR and FORP are respectively the most scalable with respect to end-to-end delay per data packet as we increase the network density.

In the absence of power control, MMBCR incurs the least energy consumed per node, closely followed by LBR. FORP incurs the maximum energy consumption per node. In terms of fairness of node usage, MMBCR is the most fair as it attempts to divert routes from nodes that have lost more battery charge towards nodes that have not lost the battery charge. FORP is the worst in terms of fairness of node usage as stable paths tend to exist for a long time and certain nodes are used more preferentially than others. LBR only manages to divert traffic from nodes that are currently part of multiple *s-d* sessions towards nodes that have been part of few s-d sessions. This strategy of LBR does not help much in shielding nodes that have lost lots of battery charge. When such nodes have not been forwarding much traffic, traffic could be diverted to such nodes. This is also the reason, why LBR has slightly larger energy consumption per node than that of MMBCR, even though LBR incurs lower energy consumption per data packet than MMBCR.

The time of first node failure for both LBR and MMBCR are high and close enough to each other, while the time of first node failure for FORP is lower. The time of first node failure for each of these routing protocols in the presence of power control, is on average, 1.4 times to that incurred in the absence of power control. For all the three routing protocols, for a given offered data traffic load, there is no significant improvement in the time of first node failure, as we double the network density, both in the presence as well as in the absence of power control. MMBCR and LBR make maximum use of power control by conserving the battery charge at the nodes while transferring data packets and use that conserved energy for route discovery at high node mobility. FORP is not much influenced by the use of power control at high node mobility as it spends most of the energy in transferring the data packets through paths of larger hop count.

The extension in the lifetime of the nodes (measured up to the $5^{th}$ node failure), after the first node failure, is least improved for MMBCR as the routing protocol attempts to equally exploit the battery charge at all the nodes in the network. LBR incurs the maximum improvement in the lifetime of the nodes beyond the time of first node failure for most of the scenarios. In certain scenarios, especially in the presence of power control, FORP sometimes incurs the maximum improvement in the lifetime of the nodes beyond the time of first node failure, mostly attributed to the unfairness of node usage.